\documentclass[apj]{emulateapj}

\usepackage{epsfig}
\usepackage{amssymb}
\usepackage{amstext}
\usepackage{amsmath}
\usepackage{amsthm}
\usepackage{verbatim}
\begin{document}
\title{An Integer Linear Programming Solution to the Telescope Network Scheduling Problem}
\author{Sotiria Lampoudi\altaffilmark{1,2}, Eric Saunders\altaffilmark{3}, Jason Eastman\altaffilmark{3}}

\altaffiltext{1}
{Earth Research Institute, 
6832 Ellison Hall, 
University of California, Santa Barbara, 
California 93106-3060, USA}

\altaffiltext{2}
{Author to whom correspondence should be addressed. 
Current Affiliation: 
Liquid Robotics Inc., 
1329 Moffet Park Dr, 
Sunnyvale, California, 94089, USA,
slampoud@gmail.com}

\altaffiltext{3}
{Las Cumbres Observatory Global Telescope,
6740 Cortona Dr,
Goleta, CA, 93117, USA}

\begin{abstract}
Telescope networks are gaining traction due to their promise of higher
resource utilization than single telescopes and as enablers of novel
astronomical observation modes. However, as telescope network sizes
increase, the possibility of scheduling them completely or even
semi-manually disappears. In an earlier paper, a step towards software
telescope scheduling was made with the specification of the
Reservation formalism, through the use of which astronomers can
express their complex observation needs and preferences. In this paper
we build on that work. We present a solution to the discretized
version of the problem of scheduling a telescope network. We derive a
solvable integer linear programming (ILP) model based on the
Reservation formalism. We show computational results verifying its
correctness, and confirm that our Gurobi-based implementation can
address problems of realistic size. Finally, we extend the ILP model
to also handle the novel observation requests that can be specified
using the more advanced Compound Reservation formalism.
\end{abstract}


\maketitle

\section{\uppercase{Introduction}}
\label{sec:introduction}

%\begin{comment}

\noindent Telescope networks have the potential to enable increased resource
utilization and novel observation modalities. Historically, observation
requests for single telescopes were made in human-readable form, and
any conflicts between requests were resolved by a person, often
working directly with the requesting astronomer in a tight feedback
loop. This type of manual scheduling is not practical for general-purpose telescope
networks containing more than a small number of telescopes, due to the
large number of competing requests received for a typical scheduling interval,
and the added complexity of choosing among multiple
resources. Further, in networks wishing to enable the study of fast
transient phenomena, manual scheduling is infeasible due to the need
for near-real-time re-scheduling responsiveness, which is necessary to
achieve these scientific objectives.

Las Cumbres Observatory Global Telescope (LCOGT) is a robotic telescope
network that currently (in September 2014) includes two 2m and nine 1m
telescopes, with plans to add a number of 0.4m telescopes in the near
future \citep{brown2013}. The telescopes are robotically controlled and connected via
the Internet to LCOGT headquarters in California. Professional
astronomers, citizen scientists and educators can apply for access to
the network on a biannual semester basis. The scientific merit of proposals is
assessed by a Time Allocation Committee (TAC), which assigns each
accepted project some amount of total time on the network and a scalar
per-unit-time priority. Each project then requests specific astronomical
observations to be conducted, not exceeding the project's total time
allocation. The motivation for the contributions of this paper is the
design and deployment of a software telescope network scheduler for
the LCOGT network \citep{saunders_spie2014}.

A software solution stack for scheduling a telescope network has three
components: (a) methods allowing the network's users to make
observation requests, (b) a scheduling algorithm capable of resolving
conflicts between users' requests to produce a viable schedule, and
(c) additional control logic that adds awareness of the state of the
network, manages schedule re-calculation (due to new input, weather,
network outages and other reasons), and deals with request
completion. In \citet{lampoudi} a formalism was developed for allowing
astronomers to express their complex observation needs and preferences
in an unambiguous, machine-readable way that would allow software to
arbitrate among them. The contribution of the present paper is a solvable
integer linear programming (ILP) model for the offline, discretized
version of the scheduling problem expressed by this formalism.

This paper is organized as follows: in section \ref{formalism} we
present the Reservation formalism that is used to communicate
astronomers' requests to the scheduler. Section \ref{ilp} presents the
solvable ILP model of the scheduling problem. Section
\ref{computational} presents computational results confirming the
correctness and evaluating the performance of the ILP
solution. Section \ref{compound} extends the model of section
\ref{ilp} to include the more complex Compound Reservations possible
in telescope networks. Related and future work are discussed in the
final section.

\section{\uppercase{The Reservation Formalism}}\label{formalism}

\noindent A request for an observation on a telescope network contains
two types of information: (a) observation-specific information about
the target of the observation, which instrument (i.e.\ camera or
spectrograph) to use, the exposure settings, etc, and (b) information
about when, where and for how long the observation can occur, based on
astronomical factors that can be calculated {\em a priori}, such as
visibility of the target during local night, lack of interference by
the moon, and so on. Information contained in (a) is necessary so that
a robotic telescope or a human operator can carry out the
observation. But it is not necessary for choosing which of many
requests to fulfill, when and where. This task, the scheduling of the
telescope network, is performed solely on the basis of the information
contained in (b). In our work the information contained in (b) is
encapsulated in a ``Reservation'': a representation of a request for
exclusive access to a resource during one contiguous chunk of time at
one or more specific times in the future.

As formally specified in \citet{lampoudi}, a Reservation $R$ is a
4-tuple $(d, p, t, W)$ where:
\begin{itemize}
\item $d$ is a scalar duration,
\item $p$ is a scalar priority,
\item $t$ is a resource (i.e.\ telescope),
\item $W$ is a list of ``windows of opportunity''
\end{itemize}

Windows of opportunity specify the times during which the observation
is possible; that is, the entire observation must fit within a single
window of opportunity.

In the vocabulary of a telescope network, a Reservation $R$ is a
request by a project with priority $p$ for exclusive access of
duration $d$ to resource $t$ during one of the windows in the list
$W$.

Typically, however, an observation can be carried out on one of many
telescopes. According to the above definition of Reservation, a
request with multiple resources (and corresponding windows of
opportunity for each resource) will result in multiple ``or''-ed
Reservations. Because this is such a common occurrence, for
compactness and with no loss of generality, we simply extend the above
definition to merge multi-resource Reservations into a single
Reservation. That is, each Reservation is now permitted to include a
list of resources, instead of a single resource, and windows of
opportunity become subscripted by resource. The resulting definition
of Reservation is the 4-tuple $(d, p, T, W)$ where:
\begin{itemize}
\item $d$ is a scalar duration,
\item $p$ is a scalar priority,
\item $T$ is a list of resources (i.e.\ telescopes), $t_i \in T$,
\item $W_i \in W$ are lists of ``windows of opportunity'', where list $W_i$
  is the list of windows corresponding to $t_i$.
\end{itemize}

\section{\uppercase{The Integer Linear Programming Model}}\label{ilp}

\noindent Given a list of Reservations we wish to find a maximum
priority subset, the subset of scheduled reservations, and an
assignment of a specific resource and start time for each scheduled
reservation, such that there are no overlaps between scheduled
reservations.

This is an offline, multi-resource, interval scheduling problem with
slack. The slack is introduced by the fact that windows can be longer
than the duration of a Reservation.

Our ILP model is inspired by a similar approach to a problem in truck
scheduling \citep{Lee}. We first discretize time into ``slots'', which
can be of uniform or non-uniform lengths. Each slot is defined by the
resource to which it corresponds and a (start time, duration) or
(start time, end time) tuple that we abbreviate as (timeslice) in the
text that follows. We then express the non-overlap constraints as
linear inequalities. Boolean variables are used to select between the
possible starting slots for each Reservation, and the sum of the
priorities of scheduled Reservations is maximised. 

The resulting ILP resembles a weighted maximum set packing problem,
which is known to be NP-complete \citep{garey}.

Specifically, our model formulation is as follows:

\subsection{Parameters}\label{variables}

\begin{itemize}
\item $I$: set of reservations
\item $T$: set of slots, each specified as a tuple: (resource, timeslice)
\item $S_i$: set of possible start slots for reservation $i$
\item $a_{ikt} = 1$ if starting reservation $i$ at $k \in S_i$ means that it will occupy slot $t$; 0 otherwise
\item $p_i$: priority of reservation $i$
\end{itemize}

\subsection{Decision variables}

$Y_{ik} = 1$ if reservation $i$ starts at $k \in S_i$; 0 otherwise

\subsection{Objectives}

Maximise the sum of the priorities of scheduled Reservations.

\begin{equation}
\max{\sum_{i \in I}\sum_{k \in S_i} p_i Y_{ik}}
\end{equation}

\subsection{Constraints}\label{constraints}

No reservation should be scheduled for more than one start.

\begin{equation}
\sum_{k \in S_i} Y_{ik} \leq 1, \forall i \in I
\end{equation}

No more than one reservation should be scheduled in each slot.
\begin{equation}
\sum_{i \in I} \sum_{k \in S_i} a_{ikt} Y_{ik} \leq 1, \forall t \in T
\end{equation}

Decision variable must be binary.
\begin{equation}
Y_{ik} \in \{0,1\}, \forall i \in I, \forall k \in S_i
\end{equation}

The inequality constraint matrix contains $|I| + |T|$ rows, i.e.\ the
sum of the number of reservations and time slots.

\section{\uppercase{Computational Results}}\label{computational}

\subsection{Correctness}

The ILP model we have just described is implemented in a software
component called the ``scheduling kernel''. This forms the core of the
software stack responsible for managing the telescope network. Our
current implementation of the kernel is in Python, and uses the Gurobi
solver \citep{gurobi}. The input to the kernel is a list of Reservation
objects which are direct implementations of the concept of a
Reservation. The kernel translates this list of Reservations into a
model description that is passed to Gurobi, and invokes the
solver. When the solver completes its run, the kernel uses the output
to modify the Reservation objects to reflect whether or not they were
scheduled, and, if they were, at what resource and starting time.

As is common software engineering practice, the kernel was unit tested
using a variety of small scheduling scenarios that can be solved
manually. But it was also desirable to validate that the kernel's
performance is what one would expect on a larger scale.

This type of validation can be achieved by using large scheduling
scenarios for which the optimal scheduling outcome is known 
{\em a  priori}, due to the way in which they were constructed. When these
scenarios are run through the scheduling kernel, it is possible to
compare the experimental performance of the kernel to this
theoretically optimal and achievable outcome.

The ``subscription rate'' of the network is defined as the ratio of
the total amount of time requested over the total amount of time
available for scheduling. The total amount of time requested is the
sum of the durations of all the reservations submitted to the
scheduler. The total amount of time available for scheduling is the
time covered by the union of all windows of opportunity.

The subscription rate is a property of the input to the scheduler. To
characterize the outcome of a scheduling run we need a corresponding
performance metric. This is the ``scheduled/requested'' (s/r) ratio,
that is the ratio of time scheduled (the sum of the durations of all
scheduled reservations) over the ratio of time requested (the sum of
the durations of all reservations that were submitted).

For subscription rates below 100\% (``undersubscribed''), it is
possible to construct problem instances for which the optimal s/r
ratio of 100\% is achievable. Given those problem instances, a
well-functioning scheduler should achieve s/r of nearly 100\%.

For subscription rates above 100\% (``oversubscribed''), it is
possible to construct problem instances for which the optimal s/r
ratio is known. Specifically, we constructed cases for which the
optimal s/r was the inverse of the subscription rate -- another way to
express that is to say that utilization was 100\%. On those problem
instances a well-functioning scheduler should achieve close to this
theoretically optimal s/r.

To produce experimental data that can be compared to the optimal
values we conducted 15 simulations of a telescope network, spanning
subscription rates between 10\% and 150\%. The size of the network was
chosen to be nine telescopes, and the time slices on all telescopes
were set to be 5 minutes long.

For each individual simulation run, i.e., for each subscription rate
value, we generated an ensemble of hundreds of reservations for which
we knew, by construction, that an optimal or close to optimal schedule
was feasible. Then the scheduling problem was made artificially harder
in two ways: first, all reservations were assigned the same 24 hour
window of opportunity; second, all were also randomly assigned to be
possible on additional resources. This had the effect of introducing
large amounts of slack and seeming contention to the problem.

%that were either uncontended (in the undersubscribed case) or for
%which the theoretical maximum s/r was feasible (in the oversubscribed
%case). All reservations were 4.9 minutes long, and not specifically
%chosen to align with slot boundaries.

\begin{figure}[!h]
  %\vspace{-0.2cm}
  \centering
   {\epsfig{file = 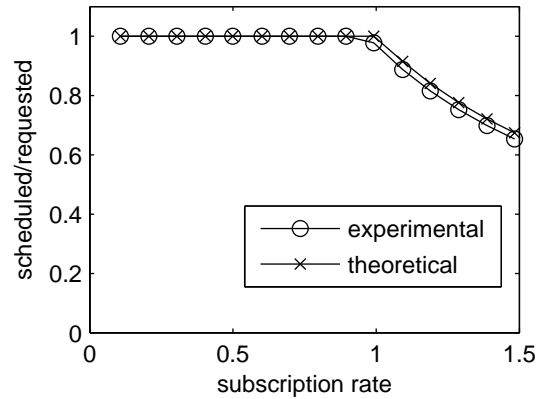, width = 7.5cm}}
  \caption{The scheduling kernel is able to achieve optimal performance in cases of undersubscription, and performs close to optimally in cases of oversubscription. The discrepancy is accounted for by the time wasted due to the discretization of time into slots.}
  \label{fig1}
\end{figure}

Figure \ref{fig1} shows that the kernel was able to match
theoretically optimal performance very closely. The small mismatch
that begins to occur around 100\% subscription can be attributed to
the small amount of time that is wasted due to the discretization of
time into slots. In the case of the artificial scenarios used for this
test, this wasted amount of time can be calculated. 

In real-world scenarios the impact of that wasted time is an open
problem. The amount of wasted time depends on (a) the distribution of
Reservation durations, and (b) the choice of slot sizes. Clearly,
smaller slots decrease the expected amount of wasted time, but
increase the size of the ILP optimization problem. We plan to study
this effect in simulation, by modeling the distribution of durations
so that we can generate appropriate synthetic workloads, and
empirically, by running the kernel on real inputs but, in
``parallel'', using different hypothetical slot sizes.

\subsection{Performance}\label{perf}

To give an idea of what currently constitutes typical and exceptional operating
conditions for the LCOGT scheduling kernel, in this section we report timings
from two real-world runs. The first is a randomly chosen typical recent run
(date: 2014-08-29). The second is the largest run that occurred since the
scheduler began official operations in May 2014 (date: 2014-06-21). They both
occurred on the same server, a 16 core Intel Xeon L5530 2.4GHz, with 24GB of
RAM. The implementation ran under Python 2.7.5 and Gurobi 5.6.2 on a Linux OS.
Gurobi was configured to use 16 threads, i.e.\ all the available cores.

In the typical run the input included 833 possible Reservations, and
seven of eleven resources were available for scheduling. This resulted
in a problem description with 20,895 rows, 138,635 columns and 479,240
nonzeros, as reported by Gurobi. This was reduced to 8,826 rows,
92,093 columns, 293,297 nonzeros by the Gurobi pre-solver. 650
Reservations were ultimately scheduled. Wall clock time spent in the
kernel (which includes the translation of the problem into a format
that can serve as an input to Gurobi, a process that we have since
further optimized) was 23.77 seconds; in the Gurobi pre-solver 6.56
seconds; in the Gurobi root relaxation routine 0.21 seconds; and in
the Gurobi integer solver 11.50 seconds. In total, 23\% of the time
was spent in kernel overhead, with the remaining time spent in Gurobi.

The biggest scheduling run during the last few months had an input of
3864 possible Reservations, roughly four times as many as the typical
run. The same fraction (7/11) of resources were available for
scheduling at the time of this run. The final schedule included 2055
of the submitted Reservations. The total kernel wall time was 76.54
seconds, of which 17.22 seconds was spent in the Gurobi pre-solver;
20.39 seconds were spent in the Gurobi root relaxation step; and 24.83
seconds were spent in the Gurobi integer solver. The kernel overhead
was thus 18\% for this larger problem. These measurements are
summarized in Table \ref{tab}.

\begin{table}[h]
\caption{Measurements for typical and largest runs.}
\label{tab} \centering
\begin{tabular}{|c|c|c|}
  \hline
   & Typical & Largest \\
  \hline
  \hline
  reservation count & 833 & 3864 \\
  \hline
   wall time (s) & 23.77 & 76.54 \\
   \hline
   \% kernel overhead & 23 & 18 \\
   \hline
   pre-solver (s) & 6.56 & 17.22 \\
   \hline
   root relaxation (s) & 0.21 & 20.39 \\
   \hline
   integer solution (s) & 11.5 & 24.83 \\
  \hline
\end{tabular}
\end{table}

\section{\uppercase{Extension to Compound Reservations}}\label{compound}

\noindent Astronomical observations requested by a particular project are
usually part of a larger scientific programme, so they are frequently
not independent of each other. When, as is most common, the
inter-dependence between observations is in the targets, instruments,
and exposures of observations, it does not affect scheduling. But it
is occasionally helpful to provide a way to express a limited form of
inter-dependence between the {\em scheduling status} of observations,
i.e.\ whether or not they were scheduled. This is useful in situations
where one of several alternative Reservations can fulfill the same
scientific objective, or when sets of Reservations must be scheduled
in an ``all-or-none'' fashion in order to be scientifically useful. We
allow for this limited type of dependency between Reservations via the
concept of a Compound Reservation, first introduced in \citet{lampoudi}.

A Compound Reservation is a set of Reservations, inter-connected by
one of two logical operators: AND and ONE-OF.

The AND operator is the traditional conjunction operator. ($r_1$ AND $r_2$)
means simply that either both reservations $r_1$ and $r_2$ should be
scheduled, or neither should be scheduled. ($r_1$ AND $r_2$ AND ... AND $r_i$)
is, by extension, defined as one would expect.

The ONE-OF operator is equivalent to a ``one-hot circuit'' in digital
circuit design. ($r_1$ ONE-OF $r_2$) means that either reservation
$r_1$ or $r_2$ should be scheduled, but not both. (Because of the
no-overlap constraint, it is possible that neither reservation can
be scheduled, making this a set packing, rather than a set
partitioning constraint.) For two arguments ONE-OF is equivalent to
XOR. The reason we use the notation ONE-OF rather than XOR is that the
implementation of a greater-than-two input XOR is not unique. The most
common implementation (i.e.\ wiring diagram) of XOR for greater than
two arguments yields a parity checker. What we need is a circuit that
evaluates to True when exactly one of its arguments is true. Since the
term ``one-hot'' is not commonly used outside digital design, we use
the more intuitive label ``ONE-OF'' for this operator.

Single-level Compound Reservations, as we describe them here, enable
some of the novel capabilities of a telescope network. They make it
possible to schedule an observation so that it occurs on one of
multiple alternative resources requiring different exposure times
(using ONE-OF), potentially increasing utilization by leveraging
flexibility. Compound Reservations also make it possible to schedule
time-series observations in an all-or-none fashion (using AND),
decreasing time wasted obtaining partial data. They make it possible
to conduct concurrent observations of a single target, or many
correlated targets, from multiple resources (using AND), which
previously required human coordination. Finally, on the LCOGT network,
which is global, they enable the tracking of stationary or moving
targets in spite of the earth's rotation, using a succession of
resources (AND), which has never before been possible. Importantly,
although an arbitrary level of Compound Reservation nesting is
conceptually possible but computationally intractable, all these
capabilities are gained using a single level of nesting, which it is
feasible to schedule.

The presence of Compound Reservations modifies the problem definition
as follows: Given a list of Reservations, where some are possibly
grouped into Compound Reservations, we wish to find a maximum priority
subset of scheduled reservations, and assign them specific resources
and start times, such that there are no overlaps between scheduled
reservations, and the constraints implied by the Compound Reservations
(single-level ANDs and ONE-OFs) are satisfied.

This extension adds the following parameters to the list of section \ref{variables}:

\begin{itemize}
\item $O$: the set of ONE-OF constraints
\item $A$: the set of AND constraints
\end{itemize}

The following constraints are added to those of section \ref{constraints}:

ONE-OF constraints.
\begin{equation}
\sum_{i \in r} \sum_{k \in S_i} Y_{ik} \leq 1, \forall r \in O_j, O_j \in O
\end{equation}

AND constraints.
\begin{equation}
\sum_{k \in S_i} Y_{ik} - \sum_{k \in S_j} Y_{jk} = 0, \forall i,j \in r, \forall r \in A_j, A_j \in A
\end{equation}

The size of the inequality constraint matrix is modified to be $|I| +
|T| + |O|$ rows, i.e.\ the sum of the number of reservations, time
slots and ONE-OF constraints.

Finally, AND constraints introduce an equality constraint matrix with
number of rows proportional to the number of reservations
participating in AND constraints. 
%, i.e.\ $\sim(\sum_{A_j \in A} |r|\in A_j)$.

\section{\uppercase{Related and Future Work}}\label{related}

The literature on ILP for interval scheduling is ubiquitous \citep[see,
e.g.][]{Schrijver, graham, potts}. Our own effort to model the telescope network
scheduling problem as an ILP problem was inspired by a similar (though more complicated, due to the presence of multiple objectives) model
for truck scheduling \citep{Lee}.

Early work in telescope scheduling, which was surveyed in
\citet{lampoudi}, was of a highly practical and heuristic nature. In
general those early approaches were difficult to evaluate for fitness
of purpose, and they commonly handled complexity, especially dynamic
volatility, through direct human intervention and decision-making.

More recently, methods adopted from the Artificial Intelligence
community, e.g.\ neural networks \citep{colome}, and from Operations
Research, e.g.\ genetic algorithms in the context of optimization
\citep{garcia}, have been making inroads in telescope network
scheduling. A necessary shift is occuring in the field towards methods
whose performance can be quantitatively compared to either theoretical
models or simulation outcomes.

For completeness, it is worth noting that there have been two previous
design iterations for the LCOGT scheduler. A randomised planning
approach was proposed in \citet{brown}. In \citet{hawkins} the
scheduling problem was broken into a hierarchy of seasonal, monthly
and adaptive planning steps, but specific implementations for those
steps were not proposed. Both of these were preliminary proposals, and
were never fully implemented or evaluated. They both reflected a
desire to steer clear of global optimisation, a stance that was
justified by citing resource availability, volatility and computational
cost. Given improvements in computational speeds, this stance was
reversed, and our present optimization approach was adopted.

Our own work is divided between, on the one hand, efforts to gain a
deeper understanding of the structure of the ILP optimization problem,
(e.g.\ analysing the structure of the conflict graph)
and, on the other hand, evaluating several practical questions
concerning the scheduling kernel and its performance. One of these is
the impact of the time discretization introduced by the ILP model, as
we explained in section \ref{perf}. Another is the choice of priority
model, or more broadly, objective functions, in order to best match
the science objectives of the network. Finally, the possibilities
implied by the compound reservation scheduling capabilities of the
kernel remain as yet uncharacterized. Such a complex feature requires
both an excellent user interface to be useful, as well as considerable
community training efforts to gain adoption. We anticipate that when
these conditions come to fruition, new statistics will become
available, and inform new models of telescope utilization, driving
forward the next iteration of development and theoretical advances.

%\end{comment}

%\vfill 
\bibliographystyle{apalike}

%{\small \bibliography{paper}}

\vfill
\end{document}